\documentclass[structabstract]{aa}

\usepackage{amsmath}   
\usepackage{graphicx}
\usepackage{txfonts}
\usepackage{natbib}
\usepackage{lscape}          

\bibpunct{(}{)}{;}{a}{}{,} 

\usepackage[pdftitle={Exomoon habitability constrained by energy flux and orbital stability}, colorlinks = true, breaklinks = true, citecolor = blue, linkcolor = blue, urlcolor = blue, pdfauthor = {Ren\'{e} Heller}]{hyperref}

\begin{document}

\title{Exomoon habitability constrained by \\ energy flux and orbital stability}


\author{R. Heller \inst{1}}

\institute{
Leibniz-Institut f\"ur Astrophysik Potsdam (AIP), An der Sternwarte 16, 14482 Potsdam, Germany, \email{rheller@aip.de}
}

\titlerunning{Exomoon habitability constrained by energy flux and orbital stability}

\authorrunning{Ren\'e Heller}

\date{Received 13 July 2012 / Accepted 30 August 2012}

\abstract
{Detecting massive satellites that orbit extrasolar planets has now become feasible, which led naturally to questions about the habitability of exomoons. In a previous study we presented constraints on the habitability of moons from stellar and planetary illumination as well as from tidal heating.}
{Here I refine our model by including the effect of eclipses on the orbit-averaged illumination. I then apply an analytic approximation for the Hill stability of a satellite to identify the range of stellar and planetary masses in which moons can be habitable. Moons in low-mass stellar systems must orbit their planet very closely to remain bounded, which puts them at risk of strong tidal heating.}
{I first describe the effect of eclipses on the stellar illumination of satellites. Then I calculate the orbit-averaged energy flux, which includes illumination from the planet and tidal heating to parametrize exomoon habitability as a function of stellar mass, planetary mass, and planet-moon orbital eccentricity. The habitability limit is defined by a scaling relation at which a moon loses its water by the runaway greenhouse process. As a working hypothesis, orbital stability is assumed if the moon's orbital period is less than $1/9$ of the planet's orbital period.}
{Due to eclipses, a satellite in a close orbit can experience a reduction in orbit-averaged stellar flux by up to about $6$\,\%. The smaller the semi-major axis and the lower the inclination of the moon's orbit, the stronger the reduction. I find a lower mass limit of $\approx~0.2\,M_\odot$ for exomoon host stars that allows a moon to receive an orbit-averaged stellar flux comparable to the Earth's, with which it can also avoid the runaway greenhouse effect. Precise estimates depend on the satellite's orbital eccentricity. Deleterious effects on exomoon habitability may occur up to $\approx~0.5\,M_\odot$ if the satellite's eccentricity is $\gtrsim~0.05$.}
{Although the traditional habitable zone lies close to low-mass stars, which allows for many transits of planet-moon binaries within a given observation cycle, resources should not be spent to trace habitable satellites around them. Gravitational perturbations by the close star, another planet, or another satellite induce eccentricities that likely make any moon uninhabitable. Estimates for individual systems require dynamical simulations that include perturbations among all bodies and tidal heating in the satellite.}


\maketitle

\section{Introduction}
\label{sec:introduction}

The detection of dozens of Super-Earths and Jupiter-mass planets in the stellar habitable zone naturally makes us wonder about the habitability of their moons \citep{1997Natur.385..234W,2010ApJ...712L.125K}. So far, no extrasolar moon has been confirmed, but dedicated surveys are underway \citep{2012ApJ...750..115K}. Several studies have addressed orbital stability of extrasolar satellite systems \citep{2010MNRAS.406.1918D,2010A&A...521A..76W} and tidal heating has been shown to be an important energy source in satellites \citep{1987AdSpR...7..125R,2006ApJ...648.1196S,2009ApJ...704.1341C}. In a recent study \citep[][HB12 in the following]{HB12}, we have extended these concepts to the illumination from the planet, i.e. stellar reflected light and thermal emission, and presented a model that invokes the runaway greenhouse effect to constrain exomoon habitability.

Earth-mass moons about Jupiter-mass planets have been shown to be dynamically stable for the lifetime of the solar system in systems where the stellar mass is greater than $>~0.15\,M_\odot$ \citep{2002ApJ...575.1087B}. I consider here whether such moons could actually be habitable. Therefore, I provide the first study of exomoon habitability that combines effects of stellar illumination, reflected stellar light from the planet, the planet's thermal emission, eclipses, and tidal heating with constraints by orbital stability and from the runaway greenhouse effect.

\section{Methods}
\label{sec:methods}

\subsection{Orbit-averaged energy flux}

To assess the habitability of a satellite, I estimated the global average
\begin{align} \label{eq:F_glob} \nonumber
\bar{F}_\mathrm{s}^\mathrm{glob} &= \frac{L_* \ (1-\alpha_\mathrm{s})}{16{\pi}a_\mathrm{*p}^2\sqrt{1-e_\mathrm{*p}^2}}
                    {\Bigg (} x_\mathrm{s} + \frac{{\pi}R_\mathrm{p}^2\alpha_\mathrm{p}}{2a_\mathrm{ps}^2} {\Bigg )} \\
             & \ \ \ \ + \frac{R_\mathrm{p}^2 \sigma_\mathrm{SB} (T_\mathrm{p}^\mathrm{eq})^4}{a_\mathrm{ps}^2} \frac{(1-\alpha_\mathrm{s})}{4} + h_\mathrm{s}
\end{align}

\noindent
of its energy flux over one stellar orbit, where $L_*$ is stellar luminosity, $a_\mathrm{*p}$ the semi-major axis of the planet's orbit about the star, $a_\mathrm{ps}$ the semi-major axis of the satellite's orbit about the planet, $e_\mathrm{*p}$ the circumstellar orbital eccentricity, $R_\mathrm{p}$ the planetary radius, $\alpha_\mathrm{p}$ and $\alpha_\mathrm{s}$ are the albedos of the planet and the satellite, respectively, $T_\mathrm{p}^\mathrm{eq}$ is the planet's thermal equilibrium temperature, $h_\mathrm{s}$ the satellite's surface-averaged tidal heating flux, $\sigma_\mathrm{SB}$ the Stefan-Boltzmann constant, and $x_\mathrm{s}$ is the fraction of the satellite's orbit that is \textit{not} spent in the shadow of the planet. Tidal heating $h_\mathrm{s}$ depends on the satellite's eccentricity $e_\mathrm{ps}$, on its radius $R_\mathrm{s}$, and strongly on $a_\mathrm{ps}$. I avoid repeating the various approaches available for a parametrization of $h_\mathrm{s}$ and refer the reader to HB12 and \citet{2011A&A...528A..27H}, where we discussed the constant-time-lag model by \citet{2010A&A...516A..64L} and the constant-phase-lag model from \citet{2008CeMDA.101..171F}. For this study, I arbitrarily chose the constant-time-lag model with a tidal time lag of the model satellites similar to that of the Earth, i.e. $\tau_\mathrm{s}~=~638$\,s \citep{1997A&A...318..975N}.

In HB12 we derived Eq.~\eqref{eq:F_glob} for $x_\mathrm{s}~=~1$, thereby neglecting the effect of eclipses on the average flux. Here, I explore the decrease of the average stellar flux on the satellite due to eclipses. Eclipses occur most frequently -- and will thus have the strongest effect on the moon's climate -- when the satellite's circum-planetary orbit is coplanar with the circumstellar orbit. {If} the two orbits are also circular and eclipses are total

\begin{equation} \label{eq:x_s}
x_\mathrm{s}~=~1-R_\mathrm{p}/({\pi}a_\mathrm{ps}) \ .
\end{equation}

\noindent
Applying the Roche criterion for a fluid-like body \citep{2010A&A...521A..76W}, I derive

\begin{equation} \label{eq:x_s_Roche}
1-\frac{1}{2.44\pi}\frac{R_\mathrm{p}}{R_\mathrm{s}} \left( \frac{M_\mathrm{s}}{M_\mathrm{p}} \right)^{1/3} < \ x_\mathrm{s} \ \leq \ 1 \ ,
\end{equation}

\noindent
where $x_\mathrm{s}~=~1$ can occur when a moon's line of nodes never crosses the planet's disk, i.e. in wide orbits. For an Earth-sized satellite about a Jupiter-sized planet Eq.~\eqref{eq:x_s_Roche} yields $x_\mathrm{s}~>~79\,\%$. Due to tidal effects, the moon's semi-major axis will typically be $>~5\,R_\mathrm{p}$, then $x_\mathrm{s}~>~1-1/({5\pi})~\approx~93.6\,\%$. In other words, eclipses will cause an orbit-averaged decrease in direct stellar illumination of $\approx~6.4\,\%$ at most in realistic scenarios. Using a constraint similar to Eq.~\eqref{eq:x_s_Roche}, \citet{2006ApJ...648.1196S} derived eclipsing effects of the same order of magnitude.

Eclipses will be most relevant for exomoons in low-mass stellar systems for two reasons. Firstly, tides raised by the star on the planet will cause the planet's obliquity to be eroded in $\ll~1$\,Gyr \citep{2011A&A...528A..27H}, and since moons will orbit their planets in the equatorial plane \citep{2011ApJ...736L..14P}, eclipses will then be most likely to occur. Secondly, exomoons in low-mass star systems must be close to their planet to ensure stability (see Sect.~\ref{sub:stability}), hence, eclipses will always be total. Below, I compute the effect of an eclipse-induced decrease of stellar irradiation for some examples.

\subsection{The circumstellar habitable zone}

The irradiation habitable zone (IHZ) is defined as the circumstellar distance range in which liquid surface water can persist on a terrestrial planet \citep{1964QB54.D63.......}. On the one hand, if a planet is too close to the star, its atmosphere will become saturated with H$_2$O. Photodissociation then drives an escape of hydrogen into space, which desiccates the planet by turning it into a runaway greenhouse \citep{1993Icar..101..108K}. On the other hand, if the planet is too far away from the star, condensation of the greenhouse gas CO$_2$ will let any liquid surface water freeze \citep{1993Icar..101..108K}. This picture can be applied to exomoons as well. However, instead of direct stellar illumination only, the star's reflected light from the planet, the planet's thermal emission, and tidal heating in the moon need to be taken into account for the global energy flux (HB12).

Analytic expressions exist that parametrize the width of the IHZ as a function of stellar luminosity. I used the set of equations from \citet{2007A&A...476.1373S} to compute the Sun-Earth equivalent distance $l_{\odot\oplus}$ between the host star and the planet -- or in this case: the moon. To a certain extent, the constraints on exomoon habitability will therefore still allow for moons with massive carbon dioxide atmospheres and strong greenhouse effect to exist in the outermost regions of the IHZ but I will not consider these special cases.

Exomoon habitability can be constrained by an upper limit for the orbit-averaged global flux. Above a certain value, typically around $300$\,W/m$^2$ for Earth-like bodies, a satellite will be subject to a runaway greenhouse effect. This value does not depend on the atmospheric composition, except that it contains water \citep{1988Icar...74..472K}. As in HB12, I used the semi-analytic expression from \citet{Pierrehumbert2012} to compute this limit, $F_\mathrm{RG}$. For computations of the satellite's radius, I used the model of \citet{2007ApJ...659.1661F} with a $68\,\%$ rock-to-mass fraction, similar to the Earth.

\subsection{Orbital stability}
\label{sub:stability}

A body's Hill radius $R_\mathrm{H}$ is the distance range out to which the body's gravity dominates the effect on a test particle. In realistic scenarios the critical semi-major axis for a satellite to remain bound to its host planet is merely a fraction of the Hill radius, i.e. $a_\mathrm{ps}~<~f~R_\mathrm{H}$ \citep{1999AJ....117..621H}, with a conservative choice for prograde satellites being $f~=~1/3$ \citep{2002ApJ...575.1087B}. If the combined mass of a planet-moon binary measured by radial-velocity is much smaller than the stellar mass, but still much larger than the satellite's mass, then $P_\mathrm{ps}/P_\mathrm{*p}~\lesssim~1/9$, where $P_\mathrm{*p}$ is the circumstellar period of the planet-moon pair and $P_\mathrm{ps}$ is the satellite's orbital period about the planet \citep{2009MNRAS.392..181K}. With $M_\mathrm{*}~\gg~M_\mathrm{p}~\gg~M_\mathrm{s}$ and the planet-moon barycenter orbiting at an orbital distance $l_\mathrm{\odot\oplus}$ to the star, Kepler's third law gives

\begin{equation} \label{eq:a_ps}
a_\mathrm{ps} \ < \ l_{\odot\oplus} \ \Bigg{[} \frac{1}{81} \frac{(M_\mathrm{p}+M_\mathrm{s})}{(M_\mathrm{*}+M_\mathrm{p})} \Bigg{]}^{1/3} \ .
\end{equation}

With this dynamical restriction in mind, we can think of scenarios in which the mass of the star is very small, so that $l_{\odot\oplus}$ lies close to it, and a moon about a planet in the IHZ must be very close to the planet to remain gravitationally bound. Then there will exist a limit (in terms of minimum stellar mass) at which the satellite needs to be so close to the planet that its tidal heating becomes so strong that it will initiate a greenhouse effect and will not be habitable. Below, I determine this minimum mass for stars to host potentially habitable exomoons.

The $P_\mathrm{ps}/P_\mathrm{*p}~\lesssim~1/9$ constraint is used as a working hypothesis, backed up by numerical studies \citep{2002ApJ...575.1087B,2006MNRAS.373.1227D,2009ApJ...704.1341C} and observations of solar system satellites. But more extreme scenarios such as retrograde, Super-Ganymede exomoons in a wide orbit about planets in the IHZ of low-mass stars are theoretically possible \citep{2010ApJ...719L.145N}. I did not consider the evolution of the planet's radius, its co-rotation radius, or the satellite's orbit about the planet. Indeed, those aspects would further reduce the likelihood of habitable satellites for given star-planet masses, because they would remove moons from the planet's orbit \citep{2002ApJ...575.1087B,2006MNRAS.373.1227D,2012ApJ...754...51S}.

\section{Results}
\label{sec:results}

\subsection{Eclipse-induced decrease of stellar illumination}

\begin{figure*}[t]
\vspace{-0.25cm}
  \centering
  \scalebox{0.75}{\includegraphics{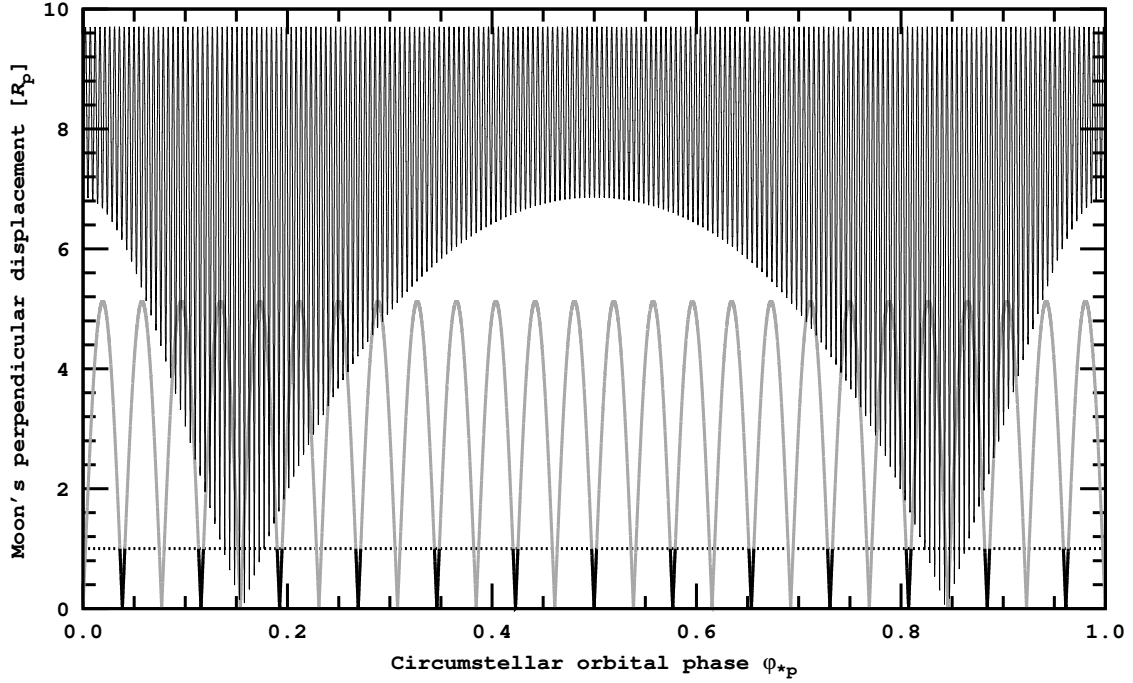}}
  \vspace{-0.4cm}
  \caption{Perpendicular displacement of a moon as seen from the star for two scenarios. The thin and highly oscillating curve corresponds to the orbit of a satellite in a Europa-wide orbit (in units of $R_\mathrm{p}$) about a Jupiter-sized planet at $1$\,AU from a Sun-like star. The satellite's orbit about the Jovian planet is inclined by $i~=~45^\circ$. The thicker gray line shows the Miranda-wide orbit of a satellite orbiting a Neptune-mass object at $\approx~0.16$\,AU from a $0.4\,M_\odot$ star, i.e. in the center of the IHZ. Here, the moon orbits in the same plane as the planet orbits the low-mass host star, i.e. $i~=~0^\circ$. Time spent in eclipse is emphasized with a thick, black line. Both orbits are normalized to the circumstellar orbital phase $\varphi_\mathrm{*p}$.}
  \label{fig:eclipses}
\end{figure*}

In Fig.~\ref{fig:eclipses} I show two examples for $x_\mathrm{s}~<~1$.\footnote{Computations were performed with my \tt{python} code \tt{exomoon.py}. Download via \url{www.aip.de/People/RHeller}.} The ordinate gives the perpendicular displacement $r_\perp$ of the satellite from the planet in units of planetary radii and as a function of circumstellar orbital phase $\varphi_\mathrm{*p}$. Thus, if $r_\perp~<~1$ and if the moon is behind the planet (and not in front of it), an eclipse occurs. Effects of the planet's penumbra are neglected.

In one scenario (thin, highly oscillating line) I placed a satellite in a Europa-wide orbit, i.e. at $9.7\,R_\mathrm{p}$, about a Jupiter-sized planet in $1$\,AU distance from a Sun-like star. I applied a stellar orbital eccentricity of $0.3$ and the moon's orbit is tilted by $45^\circ$ against the circumstellar orbit. In this specific scenario, the eccentricity causes the summer ($0~\leq~\varphi_\mathrm{*p}~\lesssim~0.18$ and $0.82~\lesssim~\varphi_\mathrm{*p}~<~1$) to be shorter than the winter ($0.18~\lesssim~\varphi_\mathrm{*p}~\lesssim~0.82$), and the inclination causes eclipses to occur only during a small part of the $365$\,d-period circumstellar orbit, namely around $\varphi_\mathrm{*p}~\approx~0.18$ and $\varphi_\mathrm{*p}~\approx~0.82$. Owing to the relatively wide satellite orbit, occultations are short compared to the moon's orbital period. Thus, effects of eclipses on the satellite's climate are small, in this case I find $x_\mathrm{s}~\approx~99.8\,\%$.

In the second scenario (gray line), I put a satellite in a close orbit at $5.1\,R_\mathrm{p}$, comparable to Miranda's orbit about Uranus, about a Neptune-mass planet, which orbits a $0.4\,M_\odot$ star in the center of the IHZ at $\approx~0.16$\,AU. The planet's circumstellar orbit with a period of roughly $36$ days is circular and the moon's three-day orbit is coplanar with the stellar orbit. In this case, transits (thick black line) occur periodically once per satellite orbit and effects on its climate will be significant with $x_\mathrm{s}~\approx~93.8\,\%$.

\subsection{Minimum stellar masses for habitable exomoons}

\begin{figure*}[t]
  \vspace{-0.05cm}
  \centering
  \scalebox{0.7}{\includegraphics{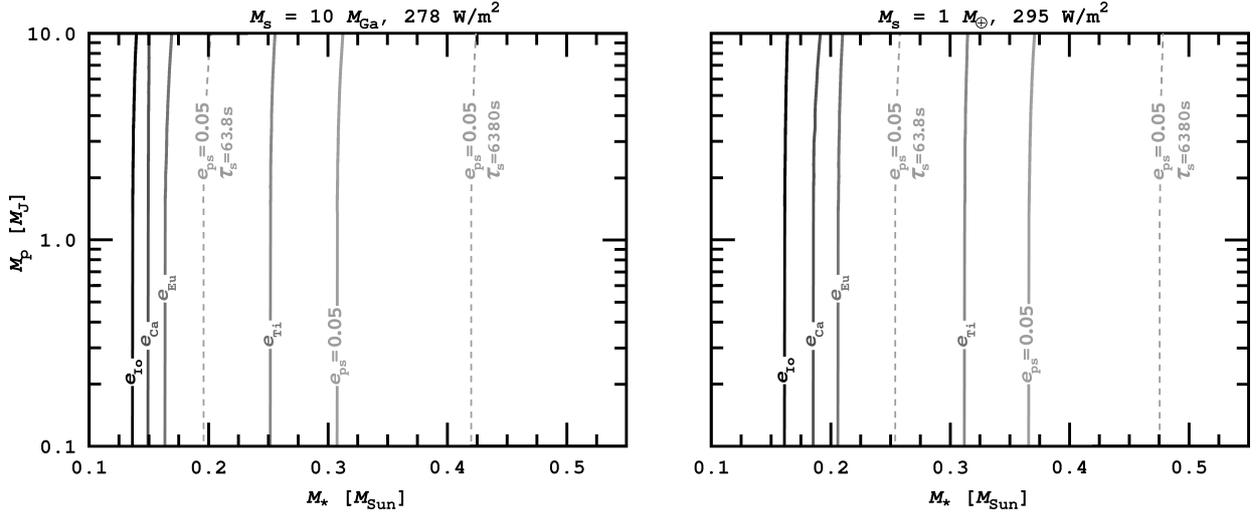}}
  \vspace{-0.2cm}
  \caption{Contours of maximum stellar masses (abscissa) for two possibly habitable moons with given planetary host masses (ordinate). Solid lines correspond to the satellite's time lag $\tau_\mathrm{s}$ of $638$\,s and one out of five eccentricities, namely those of Io, Callisto, Europa, Titan, and 0.05, from left to right. Dashed lines refer to $e_\mathrm{ps}~=~0.05$ and variation of $\tau_\mathrm{s}$ by a factor of ten. A satellite with host star and host planet masses located left to its respective eccentricity contour is uninhabitable.}
  \label{fig:exomoon_HZ}
\end{figure*}

In Fig.~\ref{fig:exomoon_HZ} contours present limiting stellar masses (abscissae) and host planetary masses (ordinate) for satellites to be habitable. I considered two moons: one with a mass ten times that of Ganymede ($10\,M_\mathrm{Ga}$, left panel) and one with the mass of the Earth ($M_\oplus$, right panel), respectively. Variation of the satellite's mass changes the critical orbit-averaged flux $F_\mathrm{RG}$ (see title of each panel). Contours of $\bar{F}_\mathrm{s}^\mathrm{glob}~=~F_\mathrm{RG}$ are drawn for five different eccentricities of the satellite, four of which are taken from the solar system: $e_\mathrm{Io}~=~0.0041$, $e_\mathrm{Ca}~=~0.0074$, $e_\mathrm{Eu}~=~0.0094$, and $e_\mathrm{Ti}~=~0.0288$ for Io, Callisto, Europa, and Titan, respectively. I also plot a threshold for a hypothetical eccentricity of $0.05$. Moons in star-planet systems with masses $M_\mathrm{*}$ and $M_\mathrm{p}$ left to a contour for a given satellite eccentricity are uninhabitable. As a confidence estimate of the tidal model, dashed contours provide the limiting mass combinations for strongly reduced ($\tau_\mathrm{s}~=~63.8$\,s) and strongly enhanced ($\tau_\mathrm{s}~=~6380$\,s) tidal dissipation in the moon for the $e_\mathrm{ps}~=~0.05$ example.

Satellites with the lowest eccentricities can be habitable even in the lowest-mass stellar systems, i.e. down to $0.1\,M_\odot$ and below for extremely circular planet-moon orbits. With increasing eccentricity, however, the satellite needs to be farther away from the planet to avoid a greenhouse effect, consequently it needs to be in a wider orbit about its planet, which in turn means that the planet needs to be farther away from the star for the moon to satisfy Eq.~\eqref{eq:a_ps}. Thus, the star needs to be more massive to have a higher luminosity and to reassure that the planet-moon binary is within the IHZ. This trend explains the increasing minimum stellar mass for increasing eccentricities in Fig.~\ref{fig:exomoon_HZ}. For $e_\mathrm{ps}~=~0.05$ this minimum stellar mass is between $0.3\,M_\odot$ for a $10\,M_\mathrm{Gan}$-mass moon (left panel) and $0.36\,M_\odot$ for an Earth-mass satellite (right panel). Uncertainties in the parametrization of tidal dissipation increase this limit to almost $0.5\,M_\odot$.

\section{Conclusions}
\label{sec:conclusions}

Eclipses of moons that are in tight orbits about their planet can substantially decrease the satellite's orbit-averaged stellar illumination. Equation~\eqref{eq:x_s} can be used to compute the reduction by total eclipses for circular and coplanar orbits. In orbits similar to the closest found in the solar system, this formula yields a reduction of about $6.4\,\%$. In orbits wider than $32$ planetary radii, the reduction is $<~1\,\%$. In tight orbits, illumination from the planet partly compensates for this reduction (HB12).

M dwarfs with masses $\lesssim~0.2\,M_\odot$ can hardly host habitable exomoons. These moons' orbits about their planet would need to be almost perfectly circularized, which means that they would need to be the only massive moon in that satellite system. Moreover, the nearby star will excite substantial eccentricities in the moon's orbit. This increases the minimum mass of host stars for habitable moons. To further constrain exomoon habitability, it will be necessary to simulate the eccentricity evolution of satellites with a model that considers both $N$-body interaction and tidal evolution. Such simulations are beyond the scope of this communication and will be conducted in a later study.

The full Kozai cycle and tidal friction model of \citet{2011ApJ...736L..14P} demonstrates very fast orbital evolution of satellites that orbit M stars in their IHZ. The star forces the satellite orbit to be eccentric, thus the moon's semi-major axis will shrink due to the tides on timescales much shorter than one Myr. Eventually, its orbit will be circular and very tight. Indeed, a moon's circum-planetary period in the IHZ about an M0 star will be even shorter than the $P_\mathrm{*p}/9$ I used as a working hypothesis \citep[see Fig.~3 in][]{2011ApJ...736L..14P}.

For low-mass stars whose IHZs are close in and whose planets' obliquities are eroded in $\ll~1$\,Gyr, eclipses will reduce the orbit-averaged stellar illumination of satellites by about a few percent. To compensate for this cooling, the planet-moon duet would need to be closer to the star, but then the moon would become uninhabitable by initiating a greenhouse effect. This contradiction makes my estimates for the minimum stellar mass for a habitable moon star more robust.

Additionally, moons about low-mass stars will most likely not experience seasons because they would orbit their planet in its equatorial plane \citep{2011ApJ...736L..14P}. The paucity of Jovian planets in the IHZ of low-mass stars \citep{2011ApJ...736...19B} further decreases the chances for habitable Earth-sized exomoons in low-mass stellar systems.

I conclude that stars with masses $\lesssim~0.2\,M_\odot$ cannot host habitable moons, and stars with masses up to $0.5\,M_\odot$ can be affected by the energy flux and orbital stability criteria combined in this paper. A model that couples gravitational scattering with tidal evolution is required to further constrain exomoon habitability about low-mass stars.

\begin{acknowledgements}
I thank the anonymous referee and the editor Tristan Guillot for their valuable comments. I am deeply grateful to Rory Barnes for our various discussions on the subject of this paper. This work has made use of NASA's Astrophysics Data System Bibliographic Services. Computations were performed with {\tt ipython 0.13} \citep{PER-GRA:2007} on {\tt python 2.7.2} and figures were prepared with {\tt gnuplot 4.4} (\url{www.gnuplot.info}) as well as with {\tt gimp 2.6} (\url{www.gimp.org}).
\end{acknowledgements}

\bibliographystyle{aa} 
\bibliography{2012-4_Exomoons_LMS}

\end{document}